\documentclass{article}
\usepackage[utf8]{inputenc}
\usepackage{graphicx}
\usepackage{subfig}
\usepackage{multicol}
\usepackage{amsmath}
\usepackage{amssymb}
\usepackage{amsthm}
\usepackage{cases}
\usepackage{hyperref}
\usepackage{enumerate}
\usepackage[letterpaper,margin=1.5in]{geometry}

\begin{document}
\title{A perturbation theory approach to the stability of the Pais-Uhlenbeck oscillator}
\author{Misael Avenda\~no-Camacho$^\dagger\mbox{}^*$, Jos\'e A. Vallejo$^\ddagger$
and Yury Vorobiev$^\dagger$ \\
{\normalsize $^\dagger\mbox{}^*$Departamento de Matem\'aticas, Universidad de Sonora (M\'exico)}\\
{\normalsize $^\ddagger$Facultad de Ciencias, Universidad Aut\'onoma de San Luis Potos\'i (M\'exico)}\\
{\normalsize $^*$CONACYT Research Fellow}\\
{\footnotesize Emails: \texttt{misaelave@mat.uson.mx,jvallejo@fc.uaslp.mx,yurimv@guaymas.uson.mx}}
}
\date{\today}
\maketitle

\begin{abstract}
We present a detailed analysis of the orbital stability of the Pais-Uhlenbeck oscillator,
using Lie-Deprit series and Hamiltonian normal form theories. In particular, we explicitly
describe the reduced phase space for this Hamiltonian system and give a proof for the existence
of stable orbits for a certain class of self-interaction, found numerically in previous works,
by using singular symplectic reduction.   
\end{abstract}

\section{Introduction}\label{sec-intro}

One of the main problems in modern field theories is their renormalizability,
that is, the possibility of canceling infinities when developing solutions to
their equations of motion as a perturbation series. In the case of quantum
theories, another (related) problem is that of unitarity. It is known since long ago that perturbatively renormalizable theories based on higher order derivatives
can be constructed, but most of these models
have been discarded in Physics because of unitarity problems in their quantization. Pais and Uhlenbeck \cite{PU50} developed 
one such model in the $50$'s,
based on a generalization of the harmonic oscillator, but taking into account 
two frequencies $w_1$, $w_2$. Their model arose in the context of gravitation
and can be described with the fourth-order differential equation
\begin{equation}\label{4thorder}
\frac{\mathrm{d}^4u}{\mathrm{d}t^4}
+(w^2_1+w^2_2)\frac{\mathrm{d}^2u}{\mathrm{d}t^2}
+w^2_1w^2_2u=0\,,
\end{equation}
for a function of time $u=u(t)$. This is the Euler-Lagrange equation for
the Lagrangian
$$
L=\left( \frac{\mathrm{d}^2u}{\mathrm{d}t^2}+w^2_1u \right)
\left( \frac{\mathrm{d}^2u}{\mathrm{d}t^2}+w^2_2u \right)\,.
$$
The Ostrogadski's second-order formalism \cite{Os50} 
generalizes the Legendre transform 
(which passes from the Lagrangian to the Hamiltonian for regular Lagrangians 
depending only on first derivatives), and allows us to find the 
corresponding Hamiltonian. If we introduce new variables
$$
\alpha =w^2_1-w^2_2,\mbox{ and }\beta =w^2_1w^2_2\,,
$$
and also
$$
\begin{cases}
q_1=\frac{1}{\sqrt{2\alpha\beta}}\left( \frac{\mathrm{d}^2u}{\mathrm{d}t^2}+w^2_1u \right) \\[5pt]
q_2=\frac{1}{\sqrt{2\alpha\beta}}\left( \frac{\mathrm{d}^2u}{\mathrm{d}t^2}+w^2_2u \right)
\end{cases}\,,
$$
the result can be written as
\begin{equation}\label{2oscillators}
H_0=\frac{1}{2}\left( p^2_1+w^2_1q^2_1\right) 
-\frac{1}{2}\left( p^2_2+w^2_1q^2_2\right)\,,
\end{equation}
where (denoting the time derivative by a point over the corresponding letter)
$$
p_1=-\sqrt{2\alpha\beta}\,\dot{u}=p_2\,.
$$
Thus, the dynamical system described by the fourth-order equation 
\eqref{4thorder}, which is a toy
model for a renormalizable theory, can be described by the Hamiltonian
\eqref{2oscillators}, which is the difference of two harmonic oscillators.
As long as the two oscillators are uncoupled, there are no physical problems
at the classical level; there exist negative energies, but these can be
interpreted in terms of the different labeling of the components, and
have the same meaning as the positive ones. But for us it is more
important to consider now the situation in which there is an interaction. If
a nonlinear interaction term is added to \eqref{2oscillators}, so the
Hamiltonian becomes, say,
\begin{equation}\label{2oscillatorsint}
H=\frac{1}{2}(p^2_1+w^2_1q^2_1)-\frac{1}{2}(p^2_2+w^2_2q^2_2) +
\frac{\Lambda}{4}(q_1+q_2)^4\,,
\end{equation} 
the interaction could lead to an exchange of energy from the
`positive' oscillator to the `negative' one, and this exchange could be done without any lower bound. An infinite amount of energy could be dissipated
by the `negative' oscillator, collapsing the system. This fact is reflected
in the existence of states of negative norm after canonical quantization
of the system, so the time evolution presents problems regarding unitarity
(although this phenomenon can be conveniently interpreted, as in \cite{IK13}, 
in order to give it a physical meaning).

The situation just described has been the main reason to discard higher-order
models as useful physical ones. However, their property of being
perturbatively renormalizable is strong enough to have prevented the complete
vanishing of interest in them, and some studies have been conducted,
numerically simulating their behavior, with the hope of finding a mechanism
that would protect unitarity or prevent the collapse of the system.
Surprisingly, what has been found is the existence of a class of
interactions (of which \eqref{2oscillatorsint} is an example) admitting
`islands of stability' (see \cite{Smi09}, \cite{IK13}
and \cite{Pav13}): For some values of the parameters
characterizing the system, the interaction generates stable periodic motions,
thus allowing these models to be considered as perfectly viable ones.
Although some heuristic arguments have been given in these papers to justify
the occurrence of these stable motions, this phenomenon continues to be
basically a finding based on numerical simulations, and its physical
origin remains unexplained. Precisely, our purpose in this note is to
show how the appearance of closed, stable orbits for a system described
by the Hamiltonian \eqref{2oscillatorsint} can be explained by using geometric and
perturbation theory and singular symplectic reduction.

The whole idea is very simple, and can be described as follows. The
complete Hamiltonian \eqref{2oscillatorsint} is regarded as a perturbation
of the Pais-Uhlenbeck oscillator \eqref{2oscillators}. The latter obviously
admits closed orbits, invariant under the action of the Hamiltonian flow
of $X_{H_0}$ (the Hamiltonian vector field of $H_0$). If the interaction term
satisfies certain conditions, seeing it as a perturbation of $H_0$ we can
think that some of the non-perturbed orbits will survive, and  we can detect them studying the fixed points of the map
expressing the returning time to a suitable Poincar\'e section. As the system
presents $U(1)$ symmetry (manifest in the periodicity of the flow of
$X_{H_0}$), it will be convenient to carry on the study of existence of 
periodic orbits in a reduced space constructed as the quotient of phase space
under the $U(1)$ action. This reduced space has the structure of a semi-algebraic
variety and, as we will see, the periodic orbits originate
precisely from its singular points (hence our use of singular reduction). These ideas have been explored in the case of
the H\'enon-Heiles system in \cite{Cus94,Cus97}, and for perturbations of the 
isotropic harmonic oscillator in \cite{Cus99}. An alternative approach to stability,
based on the notion of Lagrangian anchor, is developed in \cite{KL14,KL15}. On the other hand, M. Pav\v{s}i\v{c} has studied the stability problem under the
assumption of bounded interaction potentials. With this restriction, in \cite{ Pav13-2,Pav13-3,Pav16} he proves
 that no instabilities occur. However, in the present paper we do not restrict the potentials to the bounded case,
 so our approach is different, based on the geometric reasoning outlined above.

In the next section, we summarize the main results of the theory Hamiltonians in normal form. As we will see, the analysis is considerably simplified if
the normal form of $H$ is expressed in the so-called Hopf variables, which
for the case of the Pais-Uhlenbeck oscillator are determined in Section
\ref{sec-Hamflow}. Then, we proceed to compute the normal form (in Section
\ref{sec-normal}) and to determine the critical points of it on the reduced
phase space, which will lead to the existence of closed stable orbits in
the original system (in Section \ref{stability}). 

\section{Normal forms in perturbation theory}\label{sec-prelim}

Given a Poisson manifold $(M,P)$, with induced bracket $\{\cdot ,\cdot\}$, consider a perturbed
Hamiltonian of the form $H=H_0+\varepsilon H_1$, where $H_0$ is supposed to be integrable in general.
Hamilton's equations for $H$ are  a coupled non-linear system of differential equations,
so its solution in closed form is very difficult, or even impossible, to obtain. Perturbation
theory tries to construct approximate solutions to this system, and there are two big sets of
techniques for doing this. The first group, which comprises classical methods  such as Poincar\'e-Lindstedt or Von Zippel's, start with a known solution of the unperturbed Hamiltonian $H_0$ and add
successive corrections to it, in whose computations (usually through formal power series) enters $H_1$, and the fact that $\varepsilon$ is small. The methods in the second group, which includes the Lie-Deprit method used in this paper, try to put the system of Hamiltonian equations in an approximate, simpler form suitable to be studied by analytic tools. Thus, the perturbed Hamiltonian $H$ is said to
admit a \emph{normal form} of order $n$ if there exist a near-identity canonical transformation
on phase space such that $H$ is transformed into
\begin{equation}\label{e1}
H=\sum^n_{i=0}\varepsilon^iN_i +R_H\,,
\end{equation}
where $N_0=H_0$ and
\begin{equation}\label{e2}
\{N_i,H_0\}=0,\mbox{ for all }1\leq i\leq n\,.
\end{equation}
The function $N=\sum^n_{i=0}\varepsilon^iN_i$ is the normal form (of order $n$) of $H$.
This approach is based on the fact that whenever $\Vert H-N\Vert =\Vert R_H\Vert$ is small in a
suitable norm, the trajectories of $N$ provide us with good approximations to the true trajectories
of $H$. In particular, closed orbits for $H$ can be detected through the existence of closed orbits
for $N$.

Of course, in the setting of Poisson geometry, conditions \eqref{e1}, \eqref{e2} can be expressed in 
terms of Hamiltonian vector fields instead of functions; then, we would write respectively,
$$
X_H=\sum^n_{i=0}\varepsilon^iX_{H_i} +X_{R_H}
$$
and
$$
[X_{H_0},X_{H_i}]=0,\mbox{ for all }1\leq i \leq n\,,
$$
where, for any $F\in\mathcal{C}^\infty (M)$, $X_F=\{F,\cdot \}$ is the corresponding Hamiltonian
vector field.

To construct the desired canonical near-identity transformation that passes to the normal form,
the Lie-Deprit method resorts to generating functions. In fact, a family of canonical transformations
depending on the parameter $\varepsilon$, $x\mapsto y(x;\varepsilon)$ (where $x$ denotes
collectively the coordinates on $M$), such that $y(x;0)=x$, is defined by
\begin{equation}\label{e3}
\frac{\partial y_j}{\partial \varepsilon}=\{S,y_j\}\,,\mbox{ for }j\in\{1,\ldots ,\dim M\}\,,
\end{equation}
with $S=S(\varepsilon)$ is the generating function. Alternatively, \eqref{e3} can be written in terms
of a Lie derivative as
$$
\frac{\partial y_j}{\partial \varepsilon}=\mathcal{L}_{X_S}y_j\,,
$$
where $X_S=\{S,\cdot\}$ is the Hamiltonian vector field of $S$. It can be thought as the `$\varepsilon-$flow generator', much in the same way as $H$ is the time-flow generator.

A problem appearing in the usual formulation of the Lie-Deprit method is that, in order to compute 
$S=\sum^n_{j=0}\varepsilon^jS_j$ as a formal series (with $n$ possibly equal to $\infty$), one has
to use action--angle variables and Fourier analysis, so the formalism is of a local nature (recall
that global action--angle variables do not always exist, see \cite{Dui80,Bat88}). This
problem can be overcome in a geometric setting by exploiting the symmetries of $H_0$. In
\cite{AVV13}, a reformulation of the Lie-Deprit method is offered, valid for systems admitting
a $U(1)-$action such that the Hamiltonian vector field $X_{H_0}$ has periodic flow. When trying to 
determine the defining properties of the generating function $S$ or its Hamiltonian vector field $X_S$, 
one is lead to a set of equations called the homological equations, which basically have the form
$$
\mathcal{L}_{X_{H_0}}S_j=F_j-(j+1)N_{j+1}\,\mbox{ }j\geq 0\,,
$$
for a certain set of functions $F_j$. As previously mentioned, these equations are usually solved
by writing everything in action--angle variables, using some Fourier analysis and then averaging
over angles on orbits with constant action. The idea in \cite{AVV13} was to solve the homological
equations in a global setting, again using averaging operators, but this time constructing them
by means of geometric properties of the flow of Hamiltonian
vector fields, thus avoiding action--angle variables and the requirement that $M$ be symplectic. In
what follows we offer a brief summary of the results in \cite{AVV13}, in particular the explicit expressions for the normal forms to first and second order.

Given a manifold $M$ and a complete vector field $X\in\mathcal{X}(M)$, with periodic
flow of period function $T\in\mathcal{C}^\infty (M)$, $T>0$, we have (for any $p\in M$)
$$
\mathrm{Fl}^{t+T(p)}_X(p)=\mathrm{Fl}^t_X(p)\,,
$$
where $\mathrm{Fl}^t_X$ is the flow of $X$ evaluated at time $t\in\mathbb{R}$. In this case,
$X$ induces a $U(1)-$action by putting $(t,p)\mapsto \mathrm{Fl}^{t/w(p)}_X(p)$, where
$w=2\pi/T>0$ is the frequency function. This $U(1)-$action is periodic with constant period $2\pi$:
$$
\mathrm{Fl}^{(t+2\pi )/w(p)}_X(p)=\mathrm{Fl}^{t/w(p)}_X(p)\,.
$$
A straightforward computation shows that the generator of this $U(1)-$action is given by the vector field
$$
\Upsilon =\frac{1}{w}X\in\mathcal{X}(M)\,.
$$
Now, for any tensor field $R\in\Gamma T^r_s(M)$, its $U(1)-$averaging is defined by
$$
\left\langle R\right\rangle =\frac{1}{T}\int^T_0 (\mathrm{Fl}^t_\Upsilon )^*R\,\mathrm{d}t\,.
$$
Also, an $\mathcal{S}$ operator, mapping $\Gamma T^r_s(M)$ into itself, is defined as
$$
\mathcal{S}(R)=\frac{1}{T}\int^T_0 (t-\pi )(\mathrm{Fl}^t_\Upsilon )^*R\,\mathrm{d}t\,.
$$
In terms of these averaging operators, the solution to the homological equations (considering
now a Poisson manifold $(M,P)$ and a Hamiltonian function $H=H_0 +\varepsilon H_1$) can be 
expressed as
\begin{align*}
S_j =& \mathcal{S}\left( \frac{F_j}{w} \right)\\[6pt]
N_{j+1}=& \frac{1}{j+1}\left\langle F_j\right\rangle\,.
\end{align*}
Explicitly, the lowest order expressions for the normal forms of the perturbed Hamiltonian are
\begin{equation}\label{N1}
N_1=\left\langle H_1\right\rangle =\frac{1}{T}\int^T_0 (\mathrm{Fl}^t_{X_{H_0}})^*H_1\,
\mathrm{d}t\,,
\end{equation}
and
$$
N_2 =\frac{1}{2}\left\langle
\left\lbrace \mathcal{S}\left( \frac{H_1}{w} \right),H_1\right\rbrace
\right\rangle\,.
$$

\section{Invariants of the Hamiltonian flow of the free Pais-Uhlenbeck oscillator}\label{sec-Hamflow}

According to what we have seen in the Introduction, consider the Hamiltonian for the Pais-Uhlenbeck
oscillator in $T^\ast\mathbb{R}^2$ with coordinates $(q_1,p_1,q_2,p_2)$ and the Poisson bracket induced by the usual canonical
symplectic structure:
\begin{equation}\label{hamiltonian0}
H_0(q_1,p_1,q_2,p_2)=\frac{1}{2}(p^2_1+w^2_1q^2_1)-\frac{1}{2}(p^2_2+w^2_2q^2_2)\,.
\end{equation}
Its associated Hamiltonian vector field is readily found to be
$$
X_{H_0} = p_1\frac{\partial}{\partial q_1}-w^2_1q_1\frac{\partial}{\partial p_1}
-p_2\frac{\partial}{\partial q_2}+w^2_2q_2\frac{\partial}{\partial p_2}\,.
$$
The curves $c:I\subset\mathbb{R}\to T^\ast\mathbb{R}^2$, $c(t)=(q_1(t),p_1(t),q_2(t),p_2(t))$
which are integrals for $X_H$ satisfy the decoupled system (each dot denoting one time derivative)
$$
\begin{cases}
\ddot{q_1}+w^2_1q_1=0\\[5pt]
\ddot{q_2}+w^2_2q_2=0\,,
\end{cases}
$$
and hence, we have an action of $U(1)$ on $T^\ast\mathbb{R}^2\simeq \mathbb{R}^4$ given
by the (linear) flow of $X_{H_0}$:
\begin{align*}
\mathrm{Fl}^t_{X_{H_0}}
\begin{pmatrix} 
q_1\\[4pt]
p_1\\[4pt]
q_2\\[4pt]
p_2
\end{pmatrix}=&
\begin{pmatrix}
\frac{p_1}{w_1}\sin w_1t+q_1\cos w_1t\\[4pt]
p_1\cos w_1t-w_1q_1\sin w_1t\\[4pt]
q_2\cos w_2t-\frac{p_2}{w_2}\sin w_2t\\[4pt]
w_2q_2\sin w_2t+p_2\cos w_2t
\end{pmatrix}\,.
\end{align*}
Notice that whenever $w_1$ and $w_2$ are commensurable, this flow is periodic 
(although we may need a time rescaling to see it as an $U(1)$
action). In particular, if $w_1,w_2\in\mathbb{Z}$ are coprime, then $\mathrm{Fl}^t_{X_{H_0}}$ is
$2\pi w_ww_2-$periodic.

We are interested in determining the polynomial invariants under the action of this flow. To this
end, following \cite{Cus97}, we introduce a set of complex coordinates through the relations
$$
z_j=p_j+iw_jq_j,\quad \overline{z}_j=p_j-iw_jq_j\,,
$$
for $j\in\{1,2\}$. In terms of these, the linear flow of $X_{H_0}$ can be written in the form
$$
\mathrm{Fl}^t_{X_{H_0}}(z_1,\overline{z}_1,z_2,\overline{z}_2)=
(e^{iw_1t}z_1,e^{-iw_1t}\overline{z}_1,e^{-iw_2t}z_2,e^{iw_2t}\overline{z}_2)\,.
$$
Consider now an arbitrary monomial $z^{j_1}_1z^{j_2}_2\overline{z}^{k_1}_1\overline{z}^{k_2}_2$.
It will be invariant under the action of the Hamiltonian vector field $X_{H_0}$ if and only if
\begin{align*}
z^{j_1}_1z^{j_2}_2\overline{z}^{k_1}_1\overline{z}^{k_2}_2
=&\, (e^{iw_1t})^{j_1}(e^{-iw_2t})^{j_2}(e^{-iw_1t})^{k_1}(e^{iw_2t})^{k_2} 
z^{j_1}_1z^{j_2}_2\overline{z}^{k_1}_1\overline{z}^{k_2}_2 \\
=&\, e^{-i(-w_1j_1+w_2j_2+w_1k_1-w_2k_2)t}z^{j_1}_1z^{j_2}_2\overline{z}^{k_1}_1\overline{z}^{k_2}_2\,,
\end{align*}
that is, if and only if
$$
w_1(k_1-j_1)+w_2(j_2-k_2)=0\,.
$$
For the sake of clarity, let us call $m_1=k_1-j_1$ and $m_2=j_2-k_2$, so we must solve
\begin{equation}\label{diophantine}
m_1w_1+m_2w_2=0\,.
\end{equation}
In what follows, we will assume that $w_1,w_2\in\mathbb{Z}^+$, so we actually have a Diophantine
equation in the unknowns $(m_1,m_2)$ (the analysis of the general case is completely analogous).
Moreover, we will suppose that
$w_1$, $w_2$ are coprime; other cases, such as the resonance $w_1=w_2$, will be dealt with separately.
The condition $\mathrm{gcd}(w_1,w_2)=1$ guarantees that there exist solutions to
\eqref{diophantine}; actually a trivial one is given by $(m_1,m_2)=(-w_2,w_1)$. Notice that
$(m_1,m_2)$ is a solution if and only if $(rm_1,rm_2)$ is a solution for any $r\in\mathbb{Z}$, thus,
the set of all integer solutions (the only ones of interest for us) to \eqref{diophantine} is 
$(m_1,m_2)=(-rw_2,rw_1)$ for arbitrary $r\in\mathbb{Z}$, that is,
$$
(k_1-j_1,j_2-k_2)=(-rw_2,rw_1)\,.
$$
Let us consider the different possibilities appearing here.
\begin{enumerate}[(a)]
\item If $k_1>j_1$ and $j_2>k_2$, then $m_1,m_2>0$, and as $w_1,w_2>0$ by hypothesis, there are no solutions.
\item Analogously, if $k_1<j_1$ and $j_2<k_2$, then $m_1,m_2<0$, and as $w_1,w_2>0$, by factoring
a sign we arrive at $|m_1|w_1+|m_2|w_2=0$, which is too an equation without solutions.
\item If $k_1<j_1$ and $j_2>k_2$, then it is $r>0$, so we can write $j_1=k_1+rw_2$, $j_2=k_2+rw_1$,
and the invariant polynomial becomes
$$
z^{k_1+rw_2}_1z^{k_2+rw_1}_2\overline{z}^{k_1}_1\overline{z}^{k_2}_2
= (z_1\overline{z}_1)^{k_1}(z_2\overline{z}_2)^{k_2}(z^{w_2}_1z^{w_1}_2)^r\,.
$$
Hence, the monomials
$$
z_1\overline{z}_1,z_2\overline{z}_2,z^{w_2}_1z^{w_1}_2
$$
can be taken as generators.
\item If $k_1>j_1$ and $j_2<k_2$, then $r<0$ can be written as $r=-s$, with $s\in\mathbb{Z}^+$.
In this case, $k_1=j_1+sw_2$ and $k_2=j_2+sw_1$, so the invariant monomial is
$$
z^{j_1}_1z^{j_2}_2\overline{z}^{j_1+sw_2}_1\overline{z}^{j_2+sw_1}_2
=(z_1\overline{z}_1)^{j_1}(z_2\overline{z}_2)^{j_2}(\overline{z}^{w_2}_1\overline{z}^{w_1}_2)^s\,,
$$
and the monomials
$$
z_1\overline{z}_1,z_2\overline{z}_2,\overline{z}^{w_2}_1\overline{z}^{w_1}_2
$$
are generators.
\end{enumerate}

As a conclusion, the generators of the algebra of invariant polynomials (under the action of the
Hamiltonian flow of $X_{H_0}$) can be taken as
\begin{equation}\label{generators}
z_1\overline{z}_1,z_2\overline{z}_2,z^{w_2}_1w^{w_1}_2,\overline{z}^{w_2}_1\overline{z}^{w_1}_2\,.
\end{equation}
Alternatively, we can consider the following set of \emph{real} generators (in terms of the original
phase space variables $(q_1,q_2,p_1,p_2)$), called the \emph{Hopf variables}:
\begin{align*}
\rho_1 =& z_1\overline{z}_1=w^2_1q^2_1+p^2_1 \\
\rho_2 =& z_2\overline{z}_2=w^2_2q^2_2+p^2_2 \\
\rho_3 =& \mathrm{Re}\left( z^{w_2}_1z^{w_1}_2 \right)
	   = \mathrm{Re}\left( (p_1+iw_1q_1)^{w_2}(p_2+iw_2q_2)^{w_1} \right) \\
\rho_4 =& \mathrm{Im}\left( z^{w_2}_1z^{w_1}_2 \right) 
	   = \mathrm{Im}\left( (p_1+iw_1q_1)^{w_2}(p_2+iw_2q_2)^{w_1} \right) \,.
\end{align*}
Rather than giving the most general explicit expression, let us illustrate this result with a simple example. For instance, in the case of a $1:2$ resonance (that is, $w_1=1$, $w_2=2$), we get
\begin{align}\begin{split}\label{rhos}
\rho_1 =& q^2_1+p^2_1 \\
\rho_2 =& 4 q^2_2+p^2_2 \\
\rho_3 =& p_2(p^2_1-q^2_1)-4p_1q_1q_2 \\
\rho_4 =& 2q_2(p^2_1-q^2_1)+2q_1p_1p_2\,.
\end{split}\end{align}

There exists a certain algebraic relation satisfied by the $\rho$
variables. To begin with, we have
$$
\rho^{w_2}_1\rho^{w_1}_2=
(z_1\overline{z}_1)^{w_2}(z_2\overline{z}_2)^{w_1}
=z_1^{w_2}\overline{z}_1^{w_2}z_2^{w_1}\overline{z}_2^{w_1}\,,
$$
but this can be rearranged as
$$
z_1^{w_2}z_2^{w_1}\overline{z_1^{w_2}z_2^{w_1}}
=\mathrm{Re}^2\left( z_1^{w_2}z_2^{w_1} \right)+\mathrm{Im}^2\left( z_1^{w_2}z_2^{w_1} \right)
=\rho^2_3 + \rho^2_4\,.
$$
Thus, the real generators $(\rho_1,\rho_2,\rho_3,\rho_4)$ satisfy
\begin{equation}\label{rhoeqs}
\rho^2_3 + \rho^2_4=\rho^{w_2}_1\rho^{w_1}_2\,,\quad \rho_1,\rho_2\geq 0\,,
\end{equation}
which are the equations of a singular algebraic surface in $\mathbb{R}^4$. For the
particular case of the $1:2$ resonance, we get
\begin{equation}\label{rhoeqsbis}
\rho^2_3 + \rho^2_4=\rho^{2}_1\rho_2\,,\quad \rho_1,\rho_2\geq 0\,.
\end{equation}

Since (by a suitable rescaling) the action on $T^\ast \mathbb{R}^2\simeq \mathbb{R}^4$
of the flow of $X_{H_0}$ can be seen as a smooth $U(1)-$action, the group $U(1)$ is compact,
and the orbit space $\mathbb{R}^4/U(1)$ only contains finitely many orbit types (we will
consider the geometric structure of this orbit space in Section \ref{stability}), we can apply
the result in \cite{Sch75}, which tells us that the \emph{smooth} observables invariant under
the action of $U(1)$ are \emph{smooth} functions of the polynomial generators 
$(\rho_1,\rho_2,\rho_3,\rho_4)$.

\section{Normal form of the perturbed Hamiltonian}\label{sec-normal}

Recall from the Introduction that the main difficulty associated with the Hamiltonian 
form of the Pais-Uhlenbeck oscillator \eqref{hamiltonian0} is the following fact: When an 
interaction between the two sub-oscillators is added,
the energy can be freely transmitted from one to another. In particular, this transfer can be
done unboundedly from the oscillator $1$ to the oscillator $2$, leading to increasing negative
energies and an eventual collapse of the system\footnote{At the quantum level, this fact manifest
itself in the appearance, after canonical quantization, of states with negative norm.}.
However, it has been observed in \cite{Smi09}, \cite{IK13}
and \cite{Pav13} (among others) that for certain interactions, there are ``islands of stability'',
which have been detected numerically.

It is our intention to prove analytically the existence of such stable configurations, 
and for this we
need to write the interacting Hamiltonian in normal form.
Then, we will take the quotient by the action of the Hamiltonian flow of $H_0$ and will get
the corresponding Hamiltonian on the reduced phase space, in the next section. An important feature of this reduction process is that this reduced Hamiltonian is a function of only three among the invariant generators $(\rho_1,\rho_2,\rho_3,\rho_4)$.

For a cubic self-interacting Pais-Uhlenbeck oscillator (with the restriction $w_1\neq w_2$)
$$
\frac{\mathrm{d}^4 u}{\mathrm{d}u^4}+(w^2_1+w^2_2)\frac{\mathrm{d}^2 u}{\mathrm{d}u^2}
+w^2_1w^2_2u-\Lambda u^3 =0\,,
$$
the following Hamiltonian has been proposed in \cite{Pav13},
$$
H(q_1,p_1,q_2,p_2)=H_0+ \Lambda H_1
=\frac{1}{2}(p^2_1+w^2_1q^2_1)-\frac{1}{2}(p^2_2+w^2_2q^2_2) +
\frac{\Lambda}{4}(q_1+q_2)^4\,,
$$
where the relation between the variable $u$ and the set $(q_1,q_2,p_1,p_2)$ is obtained through
a series of substitutions and hyperbolic rotations that will not be needed here (they come from the application of Ostrogadski's second-order formalism, see \cite{Mos10}
and \cite{Pav13}). 

Due to the presence of a periodic flow, we will find it convenient to use the
techniques in \cite{AVV13}, considering $\Lambda$ as a perturbation parameter
(in fact, here $\Lambda$ is a small parameter, see \cite{Pav13}).
We begin by noticing that the Hamiltonian flow $\mathrm{Fl}^t_{X_{H_0}}$
is given by
\begin{align}\label{hamflow}
\mathrm{Fl}^t_{X_{H_0}}
\begin{pmatrix}
q_1\\[4pt]
p_1\\[4pt]
q_2\\[4pt]
p_2
\end{pmatrix}=&
\begin{pmatrix}
\frac{p_1}{w_1}\sin w_1t+q_1\cos w_1t\\[4pt]
p_1\cos w_1t-w_1q_1\sin w_1t\\[4pt]
q_2\cos w_2t-\frac{p_2}{w_2}\sin w_2t\\[4pt]
w_2q_2\sin w_2t+p_2\cos w_2t
\end{pmatrix}\,.
\end{align}
The first two components are periodic with period $T_1=\frac{2\pi}{w_2}$, while the remaining two are periodic with
$T_2=\frac{2\pi}{w_1}$. A common period $T$ is obtained by finding integers $a,b$ such that $aT_1=T=bT_2$. In
our case, $a=w_1,b=w_2$ do the job, so $T=2\pi w_1w_2$. The computation of
the second-order normal form is described in \cite{AVV13}.
In particular, the term $N_1$ is given by the averaging of the
perturbation term $H_1$ along the flow $\mathrm{Fl}^t_{X_{H_0}}$ \eqref{N1}, that is:
$$
N_1=\left\langle H_1\right\rangle = \frac{1}{2\pi w_1w_2}\int^{2\pi w_1w_2}_0 (\mathrm{Fl^t_{X_{H_0}}})^*
H_1 \,\mathrm{d}t\,.
$$

The result is
$$
N(q_1,p_1,q_2,p_2)=H_0+ \Lambda N_1 + O(\Lambda^2)
$$
where
\begin{eqnarray*}
N_1&=&
\frac{3}{32w_1^4w_2^4}\left(
w_1^4w_2^4(q^4_1+q^4_2)+
q^2_1(4w_1^4w_2^4q^2_2+ 2w_1^2w_2^4p^2_1+4w_1^4w_2^2p^2_2)\right.\\
&&\left. +q^2_2(4w_1^2w_2^4p^2_1+2w_1^4w_2^2p^2_2)+
p^2_1(w_2^4p^2_1+4w_1^2w_2^2p^2_2)+
w_1^4p^4_2.
\right)
\end{eqnarray*}
As a quick check, one can compute the Poisson brackets
$\{H_0,N_1\}=0$, as it must be for a normal form.

Now, we use \eqref{rhos} to re-express these results in terms of the Hopf invariants $\rho_i$,
obtaining
\begin{equation}\label{h0}
H_0(\rho_1,\rho_2,\rho_3,\rho_4)=\frac{1}{2}(\rho_1-\rho_2)
\end{equation}
for the free part, while
\begin{equation}\label{n1}
N_1(\rho_1,\rho_2,\rho_3,\rho_4)=\frac{3}{8}
\left(
\frac{1}{4w_1^4}\rho^2_1+\frac{1}{w_1^2w_2^2}\rho_1\rho_2+\frac{1}{4w_2^4}\rho^2_2
\right)\, .
\end{equation}
We will make use of these explicit expressions in the next section, to determine the
existence of stable periodic orbits under the flow of the Pais-Uhlenbeck oscillator.

\section{Stability analysis on the reduced phase space}\label{stability}

Recall from the discussion following \eqref{hamflow} that the Hamiltonian flow 
$\mathrm{Fl}^t_{X_{H_0}}$ is such that all its orbits are periodic with period 
$2\pi w_1w_2$. The flow of $X_{H_0}$ induces a $U(1)-$action 
on the phase space. Let $\mathcal{O}$ the orbit space given by identifying any two points of
$\mathbb{R}^4$ lying
on the same orbit of $X_{H_0}$. Any averaged vector field clearly satisfies
$\mathcal{L}_{X_{H_0}}\left\langle X\right\rangle =0$, thus leading to
$(\mathrm{Fl}^t_{X_{H_0}})^*\left\langle X\right\rangle =  \left\langle X\right\rangle$.
Therefore, the averaged vector field is completely determined
along the orbits of $X_{H_0}$ if it is known at a point. In particular, this means that
$\left\langle X\right\rangle$ descends to the orbits space.
Our reduced phase space will be obtained as a subspace of the orbits space,
by fixing a value for the momentum map of the
$U(1)-$action determined by $\mathrm{Fl}^t_{X_{H_0}}$, which
is nothing but the total energy $H_0=(\rho_1-\rho_2)/2$ (see \eqref{h0}).

At this point, there are two things to do. The first one is to geometrically identify
the reduced phase space, and the second one is to give an explicit expression for the
reduced Hamiltonian, that is, the normal form Hamiltonian $N=H_0 +\Lambda N_1 +O(\Lambda^2)$ when restricted to the reduced phase space.

\begin{figure}[h]
\centering
\subfloat{\includegraphics[trim={3cm 2cm 4cm 3cm},clip,scale=0.5]{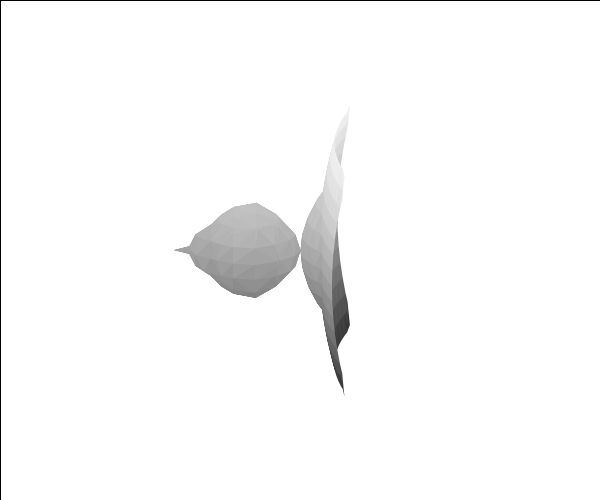}}
\subfloat{\includegraphics[trim={4cm 2cm 3cm 3.5cm},clip,scale=0.5]{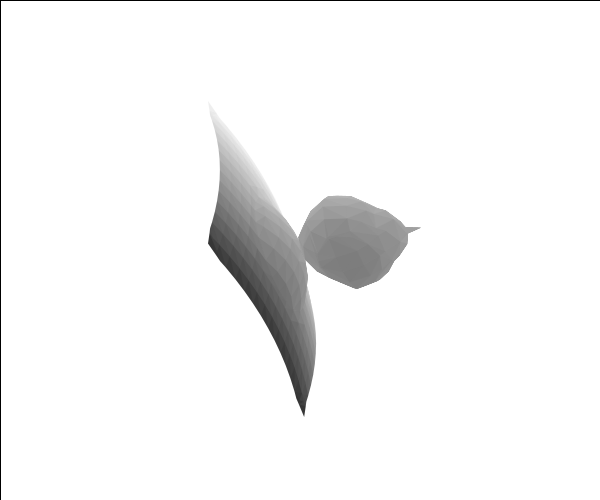}}
\caption{The reduced phase space for the PU oscillator.}
\label{fig1}
\end{figure}

For the first step, we will
follow the technique described in \cite{CKR83,Cus94} to prove that \eqref{rhoeqsbis} and the
condition of constant energy $H_0=h$, gives the algebraic description of the reduced
phase space. We use a result by Po\`enaru \cite{Poe76} (actually a corollary of the theorem by
Schwarz in \cite{Sch75}), which
states that the basic invariant polynomials separate the orbits of the Hamiltonian flow
$\mathrm{Fl}^t_{X_{H_0}}$. In our case this implies\footnote{Here we collectively denote
$(q_1,p_1,q_2,p_2)$ by $(q,p)$.} that the equality
$(\rho_1(q,p),\ldots,\rho_4(q,p))=(\rho_1(q',p'),\ldots,\rho_4(q',p'))$ holds if and only if
$(q,p)$ and $(q',p')$ belong to the same orbit. Thus, it is enough to prove that for every $(u_1,u_2,u_3,u_4)$ such that
$u^2_3+u^2_4= u^{w_2}_1u^{w_1}_2$, its inverse image under the map
$(q,p)\mapsto (\rho_1(q,p),\ldots,\rho_4(q,p))$ is precisely a single orbit of the flow
$\mathrm{Fl}^t_{X_{H_0}}$. For instance, if $u_2=0$ then $\rho_2(q,p)=0$ and necessarily
$q_2=0=p_2$ (from \eqref{rhos}). This, in turn, implies that $\rho_3=0=\rho_4$ so
we have the inverse image of $(u_1,0,0,0)$, where $u_1\geq 0$, which is the set
$\{(q_1,p_1,0,0)\in\mathbb{R}^4:q^2_1+p^2_1=u_1\}$, and this is clearly an orbit of
$\mathrm{Fl}^t_{X_{H_0}}$. The remaining cases can be done along similar lines, and
will not be written here. In what follows, we will restrict our attention to fixed negative energy values. The reduced phase space is then given by the set of equations
$$
\begin{cases}
\rho^2_3 + \rho^2_4=\rho^{w_2}_1\rho^{w_1}_2\,,\quad \rho_1,\rho_2\geq 0 \,,\\
\rho_1-\rho_2=2h\,,
\end{cases}
$$
that is,
\begin{equation}\label{algsur}
\rho^2_3 + \rho^2_4=(2h+\rho_2)^{w_2}\rho^{w_1}_2\,,\quad \rho_2\geq -2h\,.
\end{equation}
As advanced in Section \ref{sec-Hamflow}, \eqref{algsur} is the equation of a singular algebraic surface $S$
in $\mathbb{R}^3$. 
In the case of the $w_1:w_2$ resonance for the \emph{sum} of harmonic oscillators, the geometry of these
surfaces (which typically are pinched spheres) has been extensively studied by Kummer, see \cite{Kum86}.
In the present case, due to the fact that $H_0$ is an indefinite quadratic form, the surface 
\eqref{algsur} is a noncompact semialgebraic variety, as can be seen in Figure \ref{fig1} (which 
represent the above surface for $h=-1$ from two different viewpoints).

\noindent Notice that the point $(\rho_2,\rho_3,\rho_4)= (-2h, 0,0)$ is always a point on the surface (\ref{algsur}). This point is smooth when $w_2=1$, it has a conical singularity when $w_2=2$, and it has a cusp-like singularity for $w_2 \geq 3$. Therefore, the point $(-2h, 0,0)$ belongs to the reduced space if and only if $w_2=1$. This point corresponds to the curve in $\mathbb{R}^4$ given by $p_1=q_1= 0$ and $p_2+w_2^2q_2 = -2h$, called the \emph{normal mode}.
We will analyze these smoothness issues later on.

The preceding remarks lead us to consider two different cases, depending on the values of
$w_2$. Previous to the analysis of these, we need to gather some results related to Moser's 
theorem and its generalizations. The well-known theorem by Moser \cite{Mos70},
can be rephrased as follows:
Let $H=H_0+\Lambda H_1 +O(\Lambda)$ be a perturbed Hamiltonian, with $M_h$ the hypersurface
$H_0=h$. Suppose that the orbits of the Hamiltonian flow $\mathrm{Fl}^t_{X_{H_0}}$ are all periodic 
with period $T$ and let $S$ be the quotient with respect to the
induced $U(1)-$action on $M_h$. Then, to every non-degenerate critical point
$\overline{p}\in S$ of the restricted averaged perturbation
$\left. N_1\right|_S=\left. \left\langle H_1\right\rangle\right|_S$
corresponds a \emph{periodic} trajectory of the full Hamiltonian vector field $X_H$, that branches
off from the orbit represented by $\overline{p}$ and has period close to $2\pi$.
A generalization can be found in \cite{CKR83}, Theorem 6.4 there; as we will see, our setting
satisfies its hypothesis when $w_2=1$.

In order to apply these results, we must characterize the critical points of Hamiltonian vector fields in the the reduced space. First, observe that the commutator relations among generators $(\rho_1,\rho_2, \rho_3, \rho_4 )$ are given by
\begin{eqnarray}\label{relconm}
  \{\rho_1,\rho_2 \} &=& 0, \ \ \ \ \{\rho_1,\rho_3 \} = -2w_1w_2\rho_4, \ \ \ \ \{\rho_1,\rho_4 \} = 2w_1w_2\rho_3, \nonumber \\
  \{\rho_2,\rho_3 \} &=& -2w_1w_2\rho_4, \ \ \ \ \{\rho_2,\rho_4 \} = 2w_1w_2\rho_3, \nonumber\\
  \{\rho_3,\rho_4 \} &=&w_1w_2\rho_1^{w_2-1}\rho_2^{w_1-1}(w_1\rho_1+w_2\rho_2).
\end{eqnarray}
Renaming the variables $\rho_3=x$, $\rho_4=y$, and $\rho_2=z$, these relations induce a Poisson bracket on the three dimensional Euclidean space $\mathbb{R}^3=\{(x,y,z) \}$ given by
\begin{equation}\label{poibrac}
\{f, g\} = w_1w_2\left\langle\nabla g,\nabla f\times\nabla F\right\rangle\,, 
\end{equation}
where $F$ is the function
\begin{equation}\label{funcbrac}
F(x,y,z)=x^2+y^2-(z+2h)^{w_2}z^{w_1}\,,
\end{equation}
and the symbols $\left\langle\cdot ,\cdot\right\rangle$, $\times$, $\nabla$ stand for the usual inner product, cross product and nabla operator in $\mathbb{R}^3$, respectively. 
Hence, for any $f\in C^\infty(\mathbb{R}^3)$, its Hamiltonian vector field is given by
\begin{equation}\label{Hamvect}
X_f= w_1w_2 \nabla f \times \nabla F\,.
\end{equation}
It follows directly from definition \eqref{poibrac}  that the function $F(x,y,z)$ \eqref{funcbrac} 
is a Casimir of the Poisson structure \eqref{poibrac}. Thus, the symplectic leaves of the 
corresponding foliation are precisely the connected components of level sets of $F$.
If we define the mapping $P:\mathbb{R}^4 \to \mathbb{R}^4$ by
$$
P(\rho_1,\rho_2,\rho_3,\rho_4) = (\rho_3,\rho_4,\rho_2)\,,
$$
we get that $P$ is a Poisson map and $P(H_0^{-1}(h))= F^{-1}(0).$ Moreover,
\begin{equation*}
\left(P\circ \mathrm{Fl}^t_{X_{H_0}}\right)
\begin{pmatrix}
q_1\\
p_1\\
q_2\\
p_2
\end{pmatrix}
= P\begin{pmatrix}
\rho_1(p_1,q_1,p_2,q_2)\\
\rho_2(p_1,q_1,p_2,q_2)\\
\rho_3(p_1,q_1,p_2,q_2)\\
\rho_4(p_1,q_1,p_2,q_2)
\end{pmatrix}\,.
\end{equation*}
Therefore, the reduced space is contained in a symplectic leaf of $F^{-1}(0)\subset \mathbb{R}^3$ . Let us denote by $M_h$ the reduced space. Then, a realization is given by
\begin{equation}
M_h=\left\{
\begin{array}{cc}
F^{-1}(0) \text{ and } x\geq -2h & \text{ if } w_2=1\,, \\[4pt]
F^{-1}(0) \text{ and } x> -2h & \text{ if } w_2>1\,.
\end{array}
\right.
\end{equation}
Any function $f\in C^\infty(\mathbb{R}^3)$ defines a Hamiltonian vector field $\widetilde{X}_f$ 
on $M_h$ by the restriction of \eqref{Hamvect}:
\begin{equation*}
\widetilde{X}_f:= \left. (w_1w_2\nabla f\times\nabla F)\right|_{M_h}.
\end{equation*}
It also follows from \eqref{Hamvect} that the Hamiltonian vector field $\widetilde{X}_f$ has a critical point at the point $p\in M_h$ if and only if either $\nabla f(p)$ is orthogonal at $p$
to the reduced space $M_h$, or $\nabla f(p)=0$.

Next, we describe how to obtain the reduced Hamiltonian vector field corresponding to a function $G 
\in C^\infty(\mathbb{R}^4)$ such that $\{H_0, G\}=0$. As discussed in Section \ref{sec-Hamflow},  
$G$ can be expressed in terms of the Hopf variables: $G = G(\rho_1,\rho_2,\rho_3,\rho_4)$.
Writing  $\rho_1=z+2h, \rho_2=z, \rho_3=x $ and $\rho_4=x$, we obtain the function 
$Q(x,y,z)= G(z+2h,z,x,y)$. Thus, the reduced Hamiltonian vector field associated to $G$ is the 
vector field
$$
\widetilde{X}_G=\left.(  w_1w_2 \nabla Q \times \nabla F )\right|_{M_h}\,.
$$
This expression allows us to compute the critical points of the reduced vector field associated to 
the function $N_1(\rho_1,\rho_2,\rho_3,\rho_4)$ \eqref{n1}. Letting as above $K(x,y,z)=N_1(z+2h,z,x,y)$, we get
\begin{eqnarray}
K(x,y,z)(x,y,z)&=&\frac{3}{8}
\left(
\frac{1}{4w_1^4}(z+2h)^2+\frac{1}{w_1^2w_2^2}(z+2h)z+\frac{1}{4w_2^4}z^2
\right)\nonumber\\
&=&\frac{3}{8}
\left(
\frac{w_1^4+4w_1^2w_2^2+w_2^4 }{4w_1^4w_2^4}z^2+\frac{2w_1^2+w_2^2 }{w_1^4w_2^2}hz+\frac{h^2}{w_1^4}
\right)\, . \label{n1rest}
\end{eqnarray}
Hence, the reduced vector field is
\begin{equation}\label{redvector}
\widetilde{X}_{N_1}=\left.(  w_1w_2 \nabla K \times \nabla F   )\right|_{M_h}.  
\end{equation}
As we pointed out above, the critical points of \eqref{redvector} are those points $p\in M_h$ such that either $\nabla K(p)=0$ or $\nabla K(p)$ is orthogonal to $M_h$. 
By a straightforward computation, we get
\begin{equation*}
\nabla K = \left(0,0,\frac{3}{8}
\left(
\frac{w_1^4+4w_1^2w_2^2+w_2^4 }{2w_1^4w_2^4}z+\frac{2w_1^2+w_2^2 }{w_1^4w_2^2}h
\right) \right)\,.
\end{equation*}
Thus, $\nabla K(p)=0$ if and only if 
$p=(0,0,(-2h)(2w_1^2w_2^2+w_2^4)/(w_1^4+4w_1^2w_2^2+w_2^4))$. But because of 
$$
\frac{2w_1^2w_2^2+w_2^4}{w_1^4+4w_1^2w_2^2+w_2^4} <1\,,
$$
$\nabla K$ never vanishes on $M_h$.
Since, once the $z-$axis is fixed, $M_h$ is invariant under rotations on $\mathbb{R}^3$ and 
$\nabla K$ never vanishes on $M_h$, $\nabla K$ is orthogonal to $M_h$ at $(0,0,-2h)$ if and 
only if $w_2=1$ (otherwise, the point does not belong to $M_h$). We will consider the cases
$w_2=1$ and $w_2>1$ separately, as they need different treatments.

\subsection{The case \texorpdfstring{$w_2=1$}{w2=1}}
As we have seen, the point $(0,0,-2h)$ is the only critical point of the reduced Hamiltonian 
vector field $\widetilde{X}_K$ corresponding to $N_1$. A simple computation gives
\begin{equation*}
\frac{\partial F}{\partial z}(0,0,-2h)=(-2h)^{w_1}\neq 0\,.
\end{equation*}
By the implicit function theorem, there there exists a (locally defined) smooth function 
$z = g(x,y)$ such that $F(x,y,g(x,y))=0$ and $g(0,0)=-2h$. Therefore, the function $K$ in
\eqref{n1rest} has the form $\widetilde{K}=K(g(x,y))$ in a neighborhood of $(0,0,-2h)$. 
Another computation shows that
\begin{equation*}
\mathrm{Hess}(\widetilde{K}(0,0)) >0\,.
\end{equation*}
Therefore, the critical point $(0,0,-2h)$ is non-degenerate and Theorem 6.4 in \cite{CKR83} 
implies that, for small enough $\Lambda$, the Pais-Uhlenbeck oscillator has a unique stable
periodic orbit $\gamma_\Lambda$ with energy $h$ through each point $m(\Lambda)$,  sufficiently
close to $(0,0,-2h)$, with period $T(\Lambda)$, such that $H_0(m(\Lambda))\to h$ and 
$T(\Lambda) \to 2\pi w_1w_2$.

\subsection{The case \texorpdfstring{$w_2>1$}{w2>1}}
If $w_2>1$, neither $\nabla K$ vanishes on $M_h$ nor 
$\nabla K$ is orthogonal to $M_h$. $M_h$ being a quotient space, each point on it 
represents a periodic orbit of the Hamiltonian $H$. Thus, there is no point in $M_h$ whose 
periodic orbit persists as a periodic orbit of the Pais-Uhlenbeck oscillator.\\
But we must take into account that, in order to preserve smoothness, the point $(0,0,-2h)$ 
was removed  from $M_h$. 
This point precisely corresponds to the normal mode and we have seen in the previous 
case that the normal mode persists as a periodic orbit of the oscillator, so we may expect that 
bringing back $(0,0,-2h)$ could give us the orbits we seek. Unfortunately, if $w_2>1$ Moser's 
results do not apply. In this case, we will prove that the normal mode also persists as a periodic 
orbit, but we must resort to other techniques.

Let $f_2(p_1,q_1,p_2,q_2) = \frac{1}{2}(p_2+w_2^2q^2 )$. The Hamiltonian vector field
with respect to the canonical symplectic structure on $\mathbb{R}^4$, $X_{f_2}$, has 
periodic flow with periodic $T=\frac{2\pi}{w_2}$. This flow generates a free and proper
$U(1)-$action on $\mathbb{R}^2\times(\mathbb{R}^2-(0,0))$. For every fixed $h<0$, the 
level set $f_2^{-1}(-h)$ is foliated by periodic orbits of $X_{f_2}$ and a the reduced
space is given by $M_h=f_2^{-1}(-h)/U(1)$. Let us make the following change of variables
from $(p_1,q_1,p_2,q_2)$ to $(p_1,q_1,L,\theta )$:
$$
\Psi(p_1,q_1,L,\theta)=(p_1,q_1,-\sqrt{2L}\sin w_2\theta,-\frac{\sqrt{2L}}{w_2}\cos w_2 \theta),\ \ L>0\,, \ 0<\theta<\frac{2\pi}{w_2}\,.
$$
In these coordinates, the canonical symplectic form on $\mathbb{R}^2\times(\mathbb{R}^2-(0,0))$, $\mathrm{d}p_1\wedge\mathrm{d}q_1+\mathrm{d}p_2\wedge\mathrm{d}q_2$, becomes $\mathrm{d}p_1\wedge\mathrm{d}q_1+\mathrm{d}\theta\wedge\mathrm{d}L$, and the Hamiltonian
of the Pais-Uhlenbeck oscillator is
\begin{equation} \label{Hamnew}
H(p_1,q_1,L,\theta)=\frac{1}{2}(p_1^2+w_1^2q_1^2)-L + \frac{\Lambda}{4}(q_1 - \frac{\sqrt{2L}}{w_2}\cos w_2\theta)^4\,.
\end{equation}
Consider the restriction to the level set $\Sigma_h= \{(p_1,q_1,L,\theta)| L=h\}$. Since this level set is foliated by orbits of $X_{f_2}$, the Hamiltonian equations of \eqref{Hamnew} are
\begin{equation}\label{solsistnew}
\begin{cases}
\dot{\theta} =& 1 + \Lambda(q_1 -\frac{\sqrt{-2h}}{w_2}\cos w_2\theta)^3\left(\frac{\cos w_2 \theta}{w_2\sqrt{-2h}} \right)\,,\\[6pt]
\dot{p}_1 =& -w_1^2q_1-\Lambda(q_1 -\frac{\sqrt{-2h}}{w_2}\cos w_2\theta)^3\,,\\[4pt]
\dot{q}_1 =& p_1\,.
\end{cases}
\end{equation}
We now consider the cross section $\sigma_0 = \{(p_1,q_1,-h,\theta)\in \Sigma_h\,:\theta =0\}$, 
and fix the point $a=((p_1^0,q_1^0,-h,\theta^0))\in \sigma_0$. The trajectory of \eqref{solsistnew} through $a$ is:
\begin{numcases}{}
\theta(t) =t + 
\Lambda\int^t_0\left(q_1 -\frac{\sqrt{-2h}}{w_2}\cos w_2\theta\right)^3\left(\frac{\cos w_2 \theta}{w_2\sqrt{-2h}} \right)\,\mathrm{d}t\,,
\label{sistnew1}\\
p_1(t) = p_1^0 \cos w_1 t -w_1 q_1^0\sin w_1 t-\Lambda\int^t_0\left(q_1 -\frac{\sqrt{-2h}}{w_2}\cos w_2\theta\right)^3\, \mathrm{d}t\,,\label{sistnew2}\\
q_1(t) = \frac{p_1^0}{w_1}\sin w_1t + q_1\cos w_1 t\,.\label{sistnew3}
\end{numcases} 
Let $T(a,\Lambda)$ be the time elapsed between two consecutive intersections of $\sigma_0$. From equation \eqref{sistnew1}, we get
\begin{equation*}
\frac{2\pi}{w_2}=T(a,\Lambda)+\Lambda\int^{T(a,\Lambda)}_0(q_1 -\frac{\sqrt{-2h}}{w_2}\cos w_2\theta)^3\left(\frac{\cos w_2 \theta}{w_2\sqrt{-2h}} \right)\,\mathrm{d}t\,,
\end{equation*}
so $T(a, \Lambda)$ has the form 
\begin{equation}\label{time}
T(a, \Lambda)=\frac{2\pi}{w_2} +\Lambda T_1(a, \Lambda)+O(\Lambda^2)\,, 
\end{equation}
where 
$$
T_1(a)= \frac{1}{w_2\sqrt{-2h}}\int_0^{2\pi/w_2} \cos w_2t \left(\frac{\sqrt{-2h}}{w_2}\cos w_2 t - q_1^0\right)^3\,\mathrm{d}t\,.
$$
Let us remark that this is the average along the orbit of $X_{f_2}$ through $(p_1^0,q_1^0,-h,0)$. Moreover, $T_1(0,0,-h,0)\neq 0$.
Substituting (\ref{time}) in \eqref{sistnew2} and \eqref{sistnew3}, we obtain
\begin{eqnarray*}
p_1(T(a)) &=& p_1^0 +\Lambda\left(   -w_1^2 q_1^0T_1(a)-\int^{2\pi/w_2}_0(q_1^0 -\frac{\sqrt{-2h}}{w_2}\cos w_2 t)^3dt \right) +O(\Lambda^2),\\
q_1(T(a)) &=&q_1^0 +\Lambda p_1^0T_1(a)+O(\Lambda^2).
\end{eqnarray*} 
In order to prove that there exists period orbits for Pais-Uhlenbeck oscillator in $\Sigma_h$, we 
must show that, for each $\Lambda$ small enough, there exist $p_1^0(\Lambda)$ and $q_1^0(\Lambda)$ 
such that we get a fixed point:
\begin{eqnarray*}
p_1(T(p_1^0(\Lambda),q_1^0(\Lambda),-h,0,\Lambda))= p_1^0(\Lambda), \\
q_1(T(p_1^0(\Lambda),q_1^0(\Lambda),-h,0,\Lambda))= q_1^0(\Lambda).
\end{eqnarray*}
To this end, we define the following function $F: \mathbb{R}^3\rightarrow \mathbb{R}^2$,
\begin{equation*}
F\begin{pmatrix}
p_1 \\
q_1 \\
\Lambda
\end{pmatrix}=
\begin{pmatrix}
-w_1^2 q_1T_1(a)-\int^{2\pi/w_2}_0(q_1 -\frac{\sqrt{-2h}}{w_2}\cos w_2 t)^3dt  +O(\Lambda) \\
p_1T_1(a)+O(\Lambda)
\end{pmatrix}\,.
\end{equation*}
A straightforward computation shows that $F(0,0,0)=(0, 0 )^T$ and 
\begin{equation*}
\det \left( \left. \frac{\partial F}{\partial p_1\partial q_1} \right|_{(0,0,0)} \right) = \det 
\begin{pmatrix}
0 & -w_1^2 T_1(0,0,-h,0) \\
T_1(0,0,-h,0) & 0
\end{pmatrix} >0\,.
\end{equation*}
By the implicit function theorem, there exists $\delta>0$, and open neighborhood $U$ of $(0,0)$ and a function $g: (-\delta,\delta)\rightarrow U$, $g(\Lambda)=(p_1(\Lambda),q_1(\Lambda) )$ such that $g(0)=(0,0)$ and $F(g(\Lambda),\Lambda)=0$. Therefore,
\begin{eqnarray*}
p_1(T(g(\Lambda),-h,0,\Lambda))= p_1(\Lambda), \\
q_1(T(g(\Lambda),-h,0,\Lambda))= q_1(\Lambda).
\end{eqnarray*}
This fact proves that for sufficiently small $\Lambda$, the Pais-Uhlenbeck oscillator has a unique 
stable periodic orbit $\gamma_\Lambda$ with energy $h$ which branches off from the normal mode 
$\gamma$.

Summarizing, in either case, $w_2=1$ or $w_2>1$, we have that stable orbits for the self-interacting
Pais-Uhlenbeck oscillator with quartic potential exist, all of them coming from the normal mode.

\end{document}